\documentclass[aps,prl,twocolumn,showpacs,nofootinbib,amssymb,groupedaddress]{revtex4}
\usepackage{amssymb}
\usepackage{amsmath, amsthm}
\usepackage{graphicx}
\usepackage[T1]{fontenc}
\usepackage[latin1]{inputenc}
\usepackage[english]{babel}

\newcommand{\be}{\begin{equation}}
\newcommand{\ee}{\end{equation}}
\begin{document}


\title{Position and frequency shifts induced by massive modes of the
gravitational wave background in alternative gravity}


\author{Stefano Bellucci$^1$}
\email[]{bellucci@lnf.infn.it}
\author{Salvatore Capozziello$^2$}
\email[]{capozziello@na.infn.it}
\author{Mariafelicia De Laurentis$^2$}
\email[]{felicia@na.infn.it}
\author{Valerio Faraoni$^3$}
\email[]{vfaraoni@ubishops.ca}

\affiliation{$^1$ INFN Laboratori Nazionali di Frascati\\
Via Enrico Fermi 40, I-00044 Frascati, Italy
}
\affiliation{$^2$ Dip. di Scienze Fisiche,
Universit\`a di
Napoli ``Federico~II'' and INFN Sez. di Napoli\\
Compl. Universitario Monte S. Angelo, Ed.~N, Via Cinthia, I-80126
Napoli, Italy }
\affiliation{$3$ Physics Department, Bishop's University\\
Sherbrooke, Qu\'ebec, Canada J1M~1Z7 }


\begin{abstract}
Alternative theories of gravity predict the presence of massive
scalar, vector, and tensor gravitational wave modes in addition to
the standard massless spin~2 graviton of General Relativity
(GR).  The
deflection and frequency shift effects on light from distant
sources propagating through a stochastic background of
gravitational waves, containing such modes, differ  from their
counterparts in GR. Such effects are considered as
a possible signature for alternative gravity in attempts to detect
deviations from Einstein's gravity by astrophysical means.
\end{abstract}


\maketitle

\section{I. Introduction}
\setcounter{equation}{0}

Einstein's theory of GR has been tested  in its
weak-field approximation and found to pass all the available
experiments at terrestrial and Solar System scales \cite{Will}.
Outside the Solar System, the binary pulsar 1913+16
\cite{HulseTaylor} provides
indirect evidence for gravitational waves with an energy
loss  consistent
with GR. However, other theories
may produce the same change of orbital parameters:
for example, the emission of scalar radiation from
this binary within the context of scalar-tensor
gravity is necessarily small because, due to the
high symmetry of this system, the dipole moment is small. As a
consequence, the
constraints on scalar-tensor gravity imposed by the binary
pulsar are not competitive with those from Solar System tests
 (however, the binary pulsar data are sufficient to rule
out Rosen's bimetric theory) \cite{Will}.

Gravitational
lensing has provided evidence for light deflection on galactic
and cluster scales but, due to our ignorance of the detailed
mass distribution of the lens, gravitational lens systems
constitute  poor  tests for the theory of gravity (even
assuming the validity of GR, the lens
model is not unique). Instead, one
tries to  obtain information about the mass distribution in the
lens by assuming the validity of GR
and, in this context, obtains evidence for dark
matter.

No deviations from Einstein's gravity  have been detected so far
in the Solar System or binary pulsar and therefore, from the
experimental point of view, there is no compelling reason to
study alternative gravity theories. On the other hand, high energy
theories that incorporate gravity, such as superstring theory,
supergravity, and braneworld models, predict deviations from GR in
the form of extra scalar, vector, or tensor fields of
gravitational origin, massive gravitons, large extra dimensions,
higher order corrections to the Einstein equations, or violations
of the Equivalence Principle. The low-energy limit of these
theories resembles more scalar-tensor or $f(R)$ gravity than GR
\cite{bosonicstring, Stelle}. This fact, in itself, constitutes  a
motivation to explore astrophysical and other effects in
gravitational theories beyond GR. Further, the 1998 discovery
\cite{SN} that, if GR is correct, 75\% of the energy content of
the universe is in a mysterious and exotic form called dark
energy, which propels the accelerated expansion of the universe in
the present era \cite{Linderresletter}, leads one to be more
inclined towards exploring alternative theories of gravity rather
than reinforcing one's faith in Einstein's theory which, after
all, has been tested only at the post-Newtonian level and mostly
at Solar System scales. It is true that the backreaction of local
inhomogeneities in an otherwise
Friedmann-Lemaitre-Robertson-Walker universe certainly affects its
dynamics, and that this effect is obtained in pure GR without
advocating dark energy or modified gravity \cite{backreaction}.
However interesting this possibility may be, it has not been
possible to produce  evidence that the magnitude of this
backreaction effect  is such that it can explain the cosmic
acceleration observed. Backreaction, dark energy, and modified
gravity are still open possibilities, each scenario has its own
difficulties and, at present no choice between them is compelling,
and can be motivated other than by aesthetical considerations or
taste. Of course, a conservative relativist could argue that since
no deviation from GR has ever been detected, it is pointless to
actively pursue competing theories of gravity. On the other hand,
proponents of the high energy physics point of view would be
justified in replying that we may actually be detecting the first
(large scale) deviations from Einstein's theory in the cosmic
acceleration, and that it would be  foolish to ignore them.

We remind the reader that  the need to postulate dark matter in
order to explain the rotation curves at galactic and cluster
scales, has led people to doubt not only general relativity, but
even Newtonian gravity, and has produced MOND and TeVeS theories
\cite{MOND, TeVeS}, and $f(R)$ gravity with anomalous couplings to
matter \cite{extraf}. Therefore, it seems reasonable to try to
identify possible ways to test gravity beyond the Solar System.
There seem to be two conflicting  points of view, corresponding to
two different communities: deviations from Einstein's gravity are
regarded as unavoidable by  high energy physicists, their
detection being only a matter of technological limits. On the
other hand, classical relativists may regard the theories
producing such deviations as exotica, and it is certainly true
that the latter do not have experimental support so far.  While
theories of gravity alternative to general relativity are purely
hypothetical so far, they are theoretically well-motivated and
there is scope to try to detect deviations from Einstein's
gravity. It is interesting to point out that preliminary results
in positive energy theorems exist for such theories as discussed
in details in \cite{hans}. Besides, in the light of developments
in high energy physics, such possibilities should not be discarded
{\em a priori}. Eventually, experiment is the judge and the
failure to detect deviations from GR further constrains
alternative theories and worsens the fine-tuning problems that
they may have.

Here we do not want to argue in favour of GR or its competitors:
rather, we try to bridge the two points of view and we study
possible deviations from GR predicted by high energy theories.
We focus on a
possible
astrophysical effect that was studied in the past in the context
of GR, and found to be  negligible, but is
potentially interesting in alternative theories of gravity. This
effect consists of the deflection and frequency shift of a light
beam due to its propagation through a stochastic background of
gravitational waves. In many theories of gravity, extra
gravitational fields  (scalar, vector, and tensor) appear in
addition to the usual massless spin~2 graviton familiar from
GR. These modes, massless or massive, correspond
to extra degrees of freedom contained in the metric tensor
$g_{\mu\nu}$ and show up as gravitational waves emitted by early
astrophysical sources or excited by cosmological processes,
superposing to form a stochastic background. Such a background,
analogous to the cosmic microwave background of electromagnetic
waves, is well known  in GR, and the propagation
of light rays through it has been studied in detail \cite{many,
Winterberg, Zipoy, ZipoyBertotti, BertottiCatenacci, Dautcourt}.
Consider a pencil of light rays propagating from a  distant light
source (possibly at a cosmological distance) to an observer. Since
gravitational waves deflect light rays and perturb their
frequency, naively one expects  a photon undergoing $N$
scatterings in this background to be described by a random walk
and its deflections, or frequency shifts, to add stochastically as
$\sqrt{N}$. Even though the deflection is at most of linear order
in the gravitational wave amplitudes which are very small, since
the travelled distance can be large, such a cumulative (or
``$L$-'') effect that grows as $\sqrt{L}$ could compensate for it,
and it has indeed been claimed in the past \cite{Winterberg,
MarleauStarkman}. Intuition fails, however, because it is based on
familiarity with random walk processes in which the scatterers are
static or nearly static, while the massless gravitons of GR,
responsible for photon  scattering, propagate at the
speed of light. The size of the deflection (or frequency shift)
effect is a matter of {\em relative} velocities, {\em i.e.}, of
the difference between the speed of the propagating signal and
that of the perturbations from a uniform background through which
the signal propagates. When this fact is taken into account, the
cumulative $L$-effect disappears \cite{Zipoy, ZipoyBertotti,
BertottiCatenacci, Dautcourt, Linder}.  The quantitative
description of the deflection (or frequency shift) effect depends
not only on the relative speed, but also on the spin $s$ of the
field responsible for the non-stationary perturbations in the
otherwise homogeneous medium. A comprehensive quantitative
treatment is given in \cite{Linder}. The analogous situation for
massless scalar modes in scalar-tensor gravity was briefly
considered in \cite{FaraoniGunzigAA} and it was found that, in
spite of a logarithmic dependence of the rms deflection on $L$,
the effect is numerically comparable to the one in GR and,
therefore, completely negligible for practical
purposes. However, the spectrum of gravitational theories now
available is considerably larger and the consideration of
astrophysical effects due to massive fields of various spins
forming a stochastic background can potentially be of interest,
since massive fields can allow for a cumulative $L$-effect, which
will be explored in the following sections.

The plan of this paper is as follows. In Sec.~II we  briefly recall the
physics of the deflection and frequency shift effects for non-stationary
perturbations of different spins. In Sec.~III this analysis is applied to
gravitational
theories that predict deviations from GR. In Sec.~IV the
case of  modified (or
$f(R)$) gravity is studied in detail, while Sec.~V contains a
discussion and the conclusions.

\section{II. Deflections and frequency shifts caused by
propagation in a gravitational wave background}

To realize how gravitational waves induce deflections and frequency
shifts in a light ray with tangent $p^{\mu}$ that traverses them, it is
sufficient to consider
the null geodesic equation
\be
\frac{d p^{\mu}}{d\lambda} + \Gamma^{\mu}_{\alpha\beta} \, p^{\alpha}
p^{\beta}=0 \;.
\ee
By locally expanding the metric as $g_{\mu\nu}=\eta_{\mu\nu}+h_{\mu\nu}$
in an asymptotically Cartesian coordinate system, where the
perturbations
$h_{\mu\nu}$ (with $| h_{\mu\nu} |<<1$) describe gravitational waves,
computing the Christoffel symbols $\Gamma^{\mu}_{\alpha\beta}$ to first
order, and using the fact that $p^{\mu}=p_{(0)}^{\mu} +\delta
p^{\mu}=\left( 1, 0, 0, 1 \right) +\delta
p^{\mu} $ with $ \delta
p^{\mu}=\mbox{O}( h)$ for a photon with  unperturbed path along the
$z$-axis, one obtains
\begin{eqnarray}
\delta p^{\mu} & = & -\int_S^O d\lambda \, \Gamma^{\mu}_{\alpha\beta} \,
p_{(0)}^{\alpha}p_{(0)}^{\beta} \nonumber \\
&&\nonumber \\
&=&\frac{1}{2} \int_S^O dz \,
\left(
h_{00}-2h_{03}+h_{33} \right)^{, \mu} +\mbox{O}(h^2)
\;,
\end{eqnarray}
where the integral is computed along the unperturbed path from the
source
$S$ to the observer $O$. This shows that, in GR, a
gravitational wave propagating (anti)parallel to the light ray has no
effect on it, to first order.\footnote{This can be seen by adopting the
transverse-traceless gauge in which $h_{00}=h_{03}=h_{33}=0$ for a
gravitational wave propagating in the $\pm z$ direction.} If
the $h_{\mu\nu}$ describe a superposition of many waves with random
phases, directions of propagation, and polarizations, one will obtain
deflections such that $\langle \delta p^{\mu} \rangle =0$ but
$\langle \left( \delta p^{\mu} \right)^2 \rangle  \neq 0$. Therefore the
problem is whether these random deflections (for $\mu=1,2,3$) or
frequency shifts (for $\mu=0$) add stochastically. This problem has been
solved by Linder \cite{Linder} in a more general context by considering
random fluctuations due to inhomogeneities propagating with arbitrary
speed $v$ between the light source (at $z=0$) and an observer (at $z=L$)
and due to a superposition of fields of spin $s=0,1$, or $2$. By writing
the deflection due to a single mode as
\be
\theta_{\mu} =\int_0^L dz \; \epsilon_{,\mu}
\ee
and $\epsilon \left( t, \vec{x} \right)={\cal R}e\left(
\epsilon_0
\mbox{e}^{i k_{\mu}x^{\mu} } \right)$, Linder obtains the mean square
deflection
\be  \label{4}
\langle \theta^2_{\mu} \rangle = \frac{1}{2} \Sigma_{s=0}^2 \langle
{{\cal R}e}^2
\epsilon_s \rangle \sum_{n=-2}^{2s} a_n J_n \;,
\ee
where $s$ is the spin of the field responsible for the inhomogeneities,
$a_n$ are constants,  and $J_n$ are the integrals
\be
J_n=\frac{1}{\left( kL \right)^{n+1}}
\int_{\frac{-kL(1+v) }{2} }^{\frac{kL(1-v)}{2}} dy \, y^n \sin^2 y \;.
\ee
One is interested in the limit for wavenumbers $k$ and lenghts $L$ such
that $kL>>1$; in this limit the integrals $J_n$ for $n \geq 0$ cannot
cause an $L$-effect and we focus on the integrals for $n=-2, -1$,
given by \cite{Linder}
\begin{eqnarray}
J_{-1} &=& -4\left( 1+s \right) v\left( 1-v^2 \right)^s
\int_{\frac{-kL(1+v)}{2}}^{\frac{kL(1-v)}{2}} dy \, \frac{\sin^2y}{y}
\;, \\
&&\nonumber \\
J_{-2} &=& kL  \left( 1-v^2 \right)^{1+s}
\int_{\frac{-kL(1+v)}{2}}^{\frac{kL(1-v)}{2}} dy \, \frac{\sin^2y}{y^2}
\;.
\end{eqnarray}
While Linder, in the context of GR, focussed on massless
spin~2 gravitons and the limit $v\rightarrow 1$, here we are interested
in the
opposite limit for massive modes.  As shown in the next
section,
some of these modes can become very massive, corresponding to
$v\rightarrow 0$. In this case $J_{-1}$ becomes negligible and we are
left with the $J_{-2}$ contribution.

\section{III. Application to alternative theories of gravity}

In several alternative theories of gravity, massive
gravitational fields appear which can potentially give rise to an
$L$-effect. Some of them are inspected in the following.

\subsection{III.1 Scenarios with large extra dimensions}

It has been suggested \cite{ADD} that the hierarchy problem
could be solved in theories with large (sub-millimeter size) extra
spatial dimensions, in which gravitons propagate through
$(3+n)$-dimensional space  while non-gravitational physics is confined
to the ordinary three spatial dimensions (see \cite{KumarSuresh} for a
review). The $n$ extra dimensions are
compactified, {\em e.g.}, on a torus with a  radius $R_n$ and gravity
can be
strong already at the TeV scale. The gravitons propagating in the extra
dimensions acquire a mass given by
\be
m_n^2=\frac{4\pi n^2}{R_n^2} \;,
\ee
where
\be
R_n=2\cdot 10^{\frac{32-17n}{n}} \, \mbox{cm} \;.
\ee
The model is ruled out for $n=1$ and marginally ruled out for $n=2$
(for which $R_2\sim 2$~mm), but is viable for $n>2$, corresponding to
$R_n < 10^{-6}$~cm.
The dispersion relation $k_{\mu}k^{\mu}=-m_n^2 $ for the massive
gravitons yields the group velocity $v _g=\frac{c k}{\sqrt{ m^2_n
+k^2}}$.
If $m_n$ is sufficiently large, many (most) modes composing  the
gravitational wave background will have $k<<m_n$ and $v_g \sim ck/m_n <<c$. For
example,
for $n=3$, one obtains $m_3\sim 3\cdot 10^{-19/3} \; \mbox{cm}^{-1}$;
for waves of wavelength $\lambda_g \sim 10^3$~km it is $v_g\sim 10^{-2}
c$, while longer waves with $\lambda_g \sim 3 \cdot 10^8$~km$=2$~A.U.
yield  $v_g \sim 10^{-7} c$. For $n=4$ and $\lambda_g\sim 10^3 $~km, it
is $v_g\sim 10^{-17} c$.

\subsection{III.2 $f\left( R, R_{\mu\nu}R^{\mu\nu}, R_{\mu\nu\rho\sigma}
R^{\mu\nu\rho\sigma}, \Box R, \Box^2 R, \; ... \right)$ theories}

In general, in theories described by a Lagrangian density of the type
$f\left( R, R_{\mu\nu}R^{\mu\nu}, R_{\mu\nu\rho\sigma}
R^{\mu\nu\rho\sigma}, \Box R, \Box^2 R, \; ... \right)$, there are
scalar, vector, and
tensor modes, massive or massless, and these can, in principle,
contribute  to the gravitational wave background and produce an
$L$-effect. However, some of these massive modes are ghosts, which
precludes further consideration of these theories. An exception are
theories with Lagrangian of the form $f \left( R, {\cal G} \right)$,
where ${\cal G}=R^2-4R_{\mu\nu}R^{\mu\nu}
+ R_{\mu\nu\rho\sigma} R^{\mu\nu\rho\sigma} $ is the Gauss-Bonnet
combination. At least if certain conditions are satisfied, ghosts are
avoided in these theories \cite{DeFelice}.

\subsection{III.3 $N=2,8$ extended supergravity}

The supergravity multiplet in $N=2, 8$ extended supergravity contains a
graviton, a gravivector field, two Majorana gravitinos for $N=2$, and
a graviscalar
field for $N=8$. The graviscalar  violates the Weak Equivalence
Principle \cite{Scherk}, and both graviscalar and gravivector are
short-ranged. The available experiments set the  limits
 on their ranges $R_l$ and $R_{\sigma}$, respectively, \cite{Scherk,
BF}
\begin{eqnarray}
&& R_{l} \leq 0.6 \; \mbox{cm}, \;\;\;\;\;\;
 R_{l} \geq 13 \; \mbox{cm} \;\;\;\;\;\;\;(N=2), \\
&&\nonumber \\
&& R_{l} \leq 0.4 \; \mbox{cm}, \;\;\;\;\;\;
 R_{l} \geq 40 \; \mbox{m} \;\;\;\;\;\;\;(N=8), \\
&&\nonumber \\
&& R_{\sigma} \leq 0.15 \; \mbox{cm}, \;\;\;\;\;\;
60\; \mbox{m} \leq R_{\sigma} \leq 100 \; \mbox{m}.
\end{eqnarray}
If these fields are truly short-ranged, they can also contribute as
massive modes to the gravitational wave background and the
analysis of the  previous section applies.

There is scope, therefore, to consider the limit
$v\rightarrow 0$ for massive gravitons in these scenarios in the
discussion of the previous section.

\subsection{III.4 rms deflections and frequency shifts due to massive
modes}

Since $J_{-1}\rightarrow 0$ in the limit of heavy modes $v\rightarrow
0$, we are left with the contribution of
\be
J_{-2} \rightarrow kL \int_{-kL/2}^{+kL/2} dy \, \frac{\sin^2 y}{y^2}
\ee
in eq.~(\ref{4}). By using
\be
\int dy \, \frac{sin^2 y}{y^2}=\frac{ \cos (2y)}{2y} +\frac{ 2y
Si(2y)-1}{2y} \;,
\ee
where $Si(z) \equiv \int_0^z dt \, \frac{\sin t}{t}
=\frac{\pi}{2}-\int_z^{+\infty}dt\, \frac{\sin t}{t} $ is the sine
integral, one obtains
\be
J_{-2}=2\left[ \cos( kL) +kL Si( kL) -1 \right]
\ee
in the $v\rightarrow 0 $ limit.
The term $ 2\left[ \cos(kL)-1 \right]$ assumes values in the
interval $\left[ -4, 0 \right] $ and oscillates as $kL$ becomes large,
while the second term $kL
\, Si(kL)$ dominates. Since $Si(+\infty)=\pi/2$, the limit $kL>>1$
yields
the rms deflection
\be
\sqrt{ \langle \left( \theta_{\mu}\right)^2 \rangle} \approx
\sqrt{
\frac{a_{-2} \pi kL}{2}  \sum_{s=0}^2
\langle {{\cal R}e}^2 \epsilon_s \rangle }
\ee
for these modes, where an $L$-effect is indeed present and
can, in principle, compensate for small values of the gravitational wave
amplitudes $\epsilon_2$ to produce a non-negligible effect. This is not
surprising
since in the limit $v\rightarrow 0$ the propagation of the photon
reduces to a random walk.

More precisely,    keeping the dependence
of $J_{-2}$ on $v$ yields
\be
J_{-2} \simeq 2 \left( 1-v^2 \right) kL \left[ Si( kL)+ v Si(kLv)
\right]
\ee
and
\begin{eqnarray}
&& \sqrt{ \langle \theta_{\mu}^2 \rangle} = \nonumber \\
&&\nonumber \\
&& \sqrt{
\sum_{s=0}^2  2a_{-2} kL
 \left( 1-v^2 \right)  \left[ Si( kL)+ v Si(kLv)
\right] \langle {{\cal R}e}^2 \epsilon_s \rangle
} \;. \nonumber \\
&&
\end{eqnarray}

In all the scenarios listed above one can expect very massive modes for
which an $L$-effect exists and the rms deflection or frequency shift is
given, in order of magnitude,  by
\be
\sqrt{ \langle \theta_{\mu}^2\rangle} \simeq \sqrt{kL} \, \epsilon \;,
\ee
where $\epsilon$ is the magnitude of the wave amplitude for the massive
mode considered. The estimation of this quantity is difficult because it
depends on the processes generating the cosmological background, which
are subject to much speculation and large uncertainties even in
GR. The calculation of precise spectra of gravitational modes in
specific processes is beyond the purpose of this work. We assume that
detailed studies can provide, in principle, estimates of $\epsilon$
in various frequency bands following  assumptions about specific
generating processes. In this
paper we study  in more detail the case of $f(R)$ gravity.

\section{IV. Massive modes in $f(R)$ gravity}

Modified or ``$f(R)$'' gravity has been proposed recently in order to
explain the current acceleration of the universe without resorting to
dark energy \cite{CCT, CDTT, Vollick}. $f(R)$ gravity comes in three
versions: the metric \cite{CCT, CDTT},
Palatini \cite{Vollick}, and metric-affine \cite{metricaffine}
formalisms. In the
metric formalism, in
which the metric
tensor is the only independent variable and the connection is the metric
connection, the action is
\begin{equation}
\mathcal{A}=\frac{1}{2k}\int d^{4}x\sqrt{-g}f(R) +
S^{(matter)} \;,  \label{eq:2}
\end{equation}
where $f(R)$ is a non-linear function of its argument replacing
the usual Einstein-Hilbert Lagrangian  $R-2\Lambda$ (\cite{CCT,
CDTT}---see \cite{GRGrev,review, otherreviews} for  reviews).
Corrections to this Lagrangian that become important as
$R\rightarrow 0$ can explain the current acceleration of the
universe without resorting to dark energy, while early universe
physics in a strong curvature regime is instead affected by
corrections described by positive powers of $R$. Indeed, the
renormalization of GR introduces quadratic
corrections \cite{renorma}, a fact that was exploited in
Starobinsky's scenario of inflation without scalar fields
\cite{Starobinsky}. The condition $f''(R)>0$ is required in the
metric (but not in the Palatini) formalism for the absence of
tachyons \cite{DolgovKawasaki, mattmodgrav} and for non-linear
stability \cite{Frolovetc}.

Metric $f(R)$ gravity is dynamically
equivalent to an $\omega=0$ Brans-Dicke theory \cite{STequivalence,
review} with
a non-trivial potential.  In fact, by setting $\phi \equiv f'(R)$, an
equivalent
action is \cite{STequivalence, review}
\be
S=\frac{1}{2\kappa}\int d^4x \sqrt{-g} \left[ \phi R-V(\phi) \right]
+S^{(matter)} \;,
\ee
where
\be
V(\phi)=\phi R(\phi)-f(R(\phi)) \;.
\ee
The scalar degree of freedom $f'(R)$ satisfies the equation
\be
3\Box \phi +2V(\phi) -\phi\, \frac{dV}{d\phi}=\kappa\, T \;,
\ee
from which one obtains the effective mass \cite{review}
\be
m_{eff}=\sqrt{ \frac{ Rf''(R)-f'(R)}{3f''(R)} }
\ee
with $R=R(\phi)$.

In metric $f(R)$ cosmology, the dependence of the effective
mass of $\phi$ on the curvature and, therefore, on the
environmental density is exploited in the chameleon mechanism in order
to make these theories
viable. At Solar System densities, the scalar has a very short range,
thus evading the constraints imposed by Solar System and terrestrial
experiments on the equivalent Brans-Dicke theory, while at cosmological
densities this range becomes very long and can affect cosmology. This
chameleon mechanism (well-known in quintessence models
\cite{chameleon}) makes these theories viable, but at the same time
it renders the long wavelength scalar
modes forming the stochastic background effectively massless. Therefore,
the analysis of massless Brans-Dicke
scalar modes of
Ref.~\cite{FaraoniGunzigAA} applies and no $L$-effect is present.

It is more interesting, from this point of view, to consider $f(R)$
theories relevant for early universe physics.  For example, in the model
$f(R)=R+aR^2 $, it is
$m_{\phi}=1/\sqrt{6a}$. One expects the parameter $a$ weighting
quantum corrections to the Einstein-Hilbert action to be small and,
hence,
a large mass for the scalar degree of freedom $\phi$, which propagates
with group velocity $v_g \simeq ck/m_{\phi} =\sqrt{6a}\, ck$.

Assuming the conformal transformation
\begin{equation}
\widetilde{g}_{\mu\nu}=e^{2\Phi}g_{\mu\nu}\qquad \mbox{with}
\qquad e^{2\Phi}=f'(R) \;, \label{eq:3}
\end{equation}
where the prime indicates differentiation with respect to the Ricci
scalar $R$ and $\Phi$ is the ``conformal scalar field'', we obtain
the conformally equivalent Hilbert-Einstein action
\begin{equation}
\mathcal{A}
= \frac{1}{2k} \int d^{4}x \sqrt{-\widetilde{g}} \left[\widetilde{R}+
\mathcal{L}\left(\Phi\mbox{,}\Phi_{\mbox{;}\mu}\right)\right]\label{eq:4}\end{equation}
where $\mathcal{L}\left(\Phi\mbox{,}\Phi_{\mbox{;}\mu}\right)$ is
the conformal scalar field contribution derived from
\begin{equation}
\widetilde{R}_{\mu\nu}=R_{\mu\nu}+2\left(\Phi_{;\mu}\Phi_{;\nu}-
g_{\mu\nu}\Phi_{;\delta}\Phi^{;\delta}-\Phi_{;\mu\nu}-\frac{1}{2}g_{\mu\nu}\Phi^{;\delta}\,_{;\delta}\right)\label{eq:5}\end{equation}
and
\begin{equation}
\widetilde{R}=e^{-2\Phi}\left(R-6\square\Phi-6\Phi_{;\delta}\Phi^{;\delta}\right)
\;. \label{eq:6}\end{equation}
In any case, as we will see, the
$\mathcal{L}\left(\Phi\mbox{,}\Phi_{\mbox{;}\mu}\right)$-term does
not affect the  gravitational wave tensor equations  so it will not be
considered further.\footnote{Actually, a scalar component of
gravitational
radiation is often considered \cite{Maggiore,capozzcorda}, but here we
are taking into account only the genuine tensor part of the stochastic
background.}

Beginning with the action (\ref{eq:4}) and deriving the
Einstein-like
conformal equations, the gravitational wave equations expressed in the conformal
metric $\widetilde{g}_{\mu\nu}$ are
\begin{equation}
\widetilde{\square} \, \widetilde{h}_{i}^{j}=0\label{eq:7} \;.
\end{equation}
Since
no scalar perturbation couples to the tensor part of the gravitational
waves,  we have
$\delta\Phi=0$ and then
\begin{equation}
\widetilde{h}_{i}^{j}=\widetilde{g}^{lj}\delta\widetilde{g}_{il}=e^{-2\Phi}g^{lj}e^{2\Phi}\delta
g_{il}=h_{i}^{j}\label{eq:8}
\end{equation}
which means that $h_{i}^{j}$ is a conformal invariant. As a consequence,
the plane-wave amplitude defined by
$h_i^j (t, x)=h(t) \, e_{i}^{j} \exp(ik_{l}x^{l}),$ where $e_{i}^{j}$ is
the
polarization tensor, are the same in both metrics. In any case,
the d'Alembert operator transforms as \begin{equation}
\widetilde{\square}=e^{-2\Phi}\left(\square+2\Phi^{;
\lambda}\nabla_{\lambda}\right)\label{eq:9}\end{equation}
and this means that the background is changing while the tensor
wave amplitude is not.

In order to study the cosmological stochastic background, the
operator (\ref{eq:9}) can be specified for a
Friedmann-Robertson-Walker (FRW) metric  given by
\be
ds^2=-dt^2+a^2(t)\left( dx^2+dy^2+dz^2 \right) \;,
\ee
and then eq.~(\ref{eq:7})
becomes
\begin{equation}
\ddot{h}+\left(3H+2\dot{\Phi}\right)\dot{h}+k^{2}a^{-2}h=0
\;, \label{eq:10}
\end{equation}
where ${\displaystyle \square=\frac{\partial^2 }{\partial
t^{2}}+3H\frac{\partial}{\partial t}}$
and $k$ is the wave number.

It is worth stressing that eq.~(\ref{eq:10}) applies to any
$f(R)$ theory whose conformal transformation can be defined as
${\displaystyle e^{2\Phi}=f'(R).}$ The solution, {\em i.e.}, the
gravitational wave
amplitude, depends on the specific cosmological background ({\em i.e.},
$a(t)$) and the specific theory of gravity ({\em i.e.}, $\Phi(t)$)
\cite{SCF1}.
Considering also the conformal time
$d\eta=dt/a$, eq.~(\ref{eq:10}) reads
\begin{equation}
\frac{d^2 h}{d\eta^2}+\frac{2}{\chi}\, \frac{d\chi}{d\eta}
\frac{dh}{d\eta}+k^{2}h=0  \;, \label{eq:16}
\end{equation}
where $\chi \equiv a \, \mbox{e}^{\Phi}$.
Inflation means that $a(t)=a_{0}\exp(Ht)$ and then $\eta=\int dt/
a=(aH)^{-1}$ and $\frac{d\chi}{\chi d\eta}=-\eta^{-1}$. The exact
solution of
(\ref{eq:16}) is
\begin{equation}
 h(\eta)=\sqrt{2} \, k^{-2}
 \left[ C_{1}\sin k\eta+ C_{2}\cos k\eta\right] \;. \label{eq:17}
\end{equation}
Inside the $H^{-1}$ radius we have $k\eta\gg 1.$ Furthermore,
considering the absence of gravitons in the initial vacuum state,
we have only negative-frequency modes and then the adiabatic
behavior is
\begin{equation}
h=  \sqrt{\frac{2}{\pi}} \,
k^{1/2}\frac{1}{aH} \, C\exp(-ik\eta)\,.\label{eq:18}
\end{equation}

At the first horizon crossing  $(aH=k)$, the averaged amplitude
of the perturbation  $A_{h}=(k/2\pi)^{3/2}\left|h\right|$ is
\begin{equation}
A_{h}=\frac{C}{2\pi^{2}} \label{eq:19} \;.
\end{equation}
When the scale $a/k$ grows larger than the Hubble radius $H^{-1}$,
the growing  mode of evolution is frozen, that is, it is constant.
This situation corresponds to the limit $k\eta\ll 1$ in
eq.~(\ref{eq:17}). Since $\Phi$ acts as the inflaton field, it is
$\Phi\sim 0$ at re-entry after the end of inflation. Then the
amplitude $A_{h}$ of the wave is preserved until the second
horizon crossing after which it can be observed, in principle, as
an anisotropy perturbation in the cosmic microwave background. It
can be shown that $\bigtriangleup T/T\lesssim A_{h}$ as an upper
limit to $A_{h}$ since other effects can contribute to the
background anisotropy \cite{staro}. From these considerations, it
is clear that the only relevant quantity is the initial amplitude
$C$ in eq.~(\ref{eq:18}), which is conserved until re-entry into
the horizon. Such an amplitude directly depends on the fundamental
mechanism generating the perturbations. Inflation gives rise to
processes capable of producing perturbations as zero-point energy
fluctuations. Such a mechanism depends on the  theory of
gravitation adopted and then $(\bigtriangleup T/T)$ could
constitute a further constraint to select a suitable
$f(R)$-theory. Considering a single graviton in the form of a
monochromatic wave, its zero-point amplitude is derived through
the equal time commutation relations
\begin{equation}
\left[h(t,x),\,\pi_{h}(t,y)\right]=i \, \delta^{3}(x-y) \;, \label{eq:20}
\end{equation}
where the amplitude $h$ is the
field and $\pi_{h}$ is the conjugate momentum operator. Writing
the Lagrangian for $h$
\begin{equation}
\widetilde{\mathcal{L}}=\frac{1}{2}\sqrt{-\widetilde{g}}
\, \widetilde{g}^{\mu\nu}h_{;\mu}h{}_{;\nu}\label{eq:21}
\end{equation}
in the conformal FRW metric $\widetilde{g}_{\mu\nu}$  ($h$ is
conformally invariant), we obtain
\begin{equation}
\pi_{h}=\frac{\partial\widetilde{\mathcal{L}}}{\partial\dot{h}}
=e^{2\Phi}a^{3}\dot{h} \;.\label{eq:22}\end{equation}

Then, eq.~(\ref{eq:20}) becomes
\begin{equation}
\left[h(t,x),\,\dot{h}(t,y)\right]=i
\, \frac{\delta^{3}(x-y)}{a^{3}e^{2\Phi}}\label{eq:23}
\end{equation}
and the fields $h$ and $\dot{h}$ can be expanded in terms of
creation and annihilation operators
\begin{equation}
h(t,x)=\frac{1}{(2\pi)^{3/2}}\int d^{3}k\left[h(t)e^{-ikx}+h^{*}(t)e^{+ikx}\right],\label{eq:24}
\end{equation}
\begin{equation}
\dot{h}(t,x)=\frac{1}{(2\pi)^{3/2}}\int d^{3}k\left[\dot{h}(t)e^{-ikx}+\dot{h}^{*}(t)e^{+ikx}\right].\label{eq:25}
\end{equation}

\begin{figure}[ht]
\includegraphics[scale=0.5]{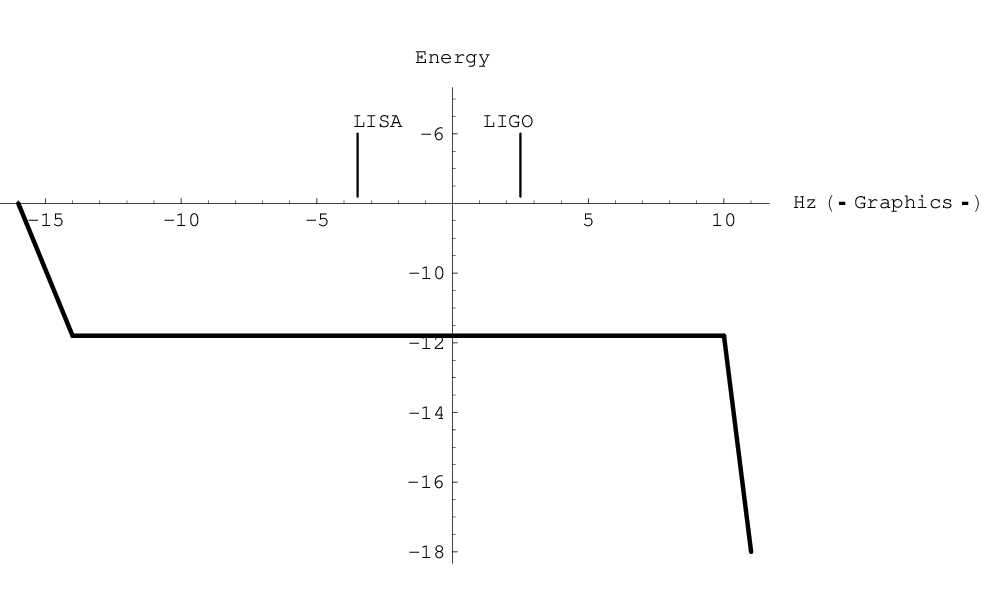}
\caption{The spectrum of relic scalar gravitational waves  in
inflationary models
is flat over a wide range of frequencies. The horizontal axis is
$\log_{10}$ of frequency, in Hz. The vertical axis is
$\log_{10}\Omega_{gsw}$. The inflationary spectrum rises quickly
at low frequencies (wave which re-entered the Hubble sphere
after the universe became matter-dominated) and falls off above
the (appropriately redshifted) frequency scale $f_{max}$
associated with the fastest characteristic time of the phase
transition at the end of inflation. The amplitude of the flat
region depends only on the energy density during the inflationary
stage; we have chosen the largest amplitude consistent with the
{\em WMAP} constraints on scalar perturbations. This means that, at {\em
LIGO}
and {\em LISA} frequencies, $\Omega_{sgw}<2.3 \cdot 10^{-12}$.}
\label{fig:spectrum}
\end{figure}

The commutation relations in conformal time are then
\begin{equation}
\left[hh'^{*}-h^{*}h'\right]=\frac{i(2\pi)^{3}}{a^{3} \mbox{e}^{2\Phi}}
\;. \label{eq:26}
\end{equation}
The substitution of eqs.~(\ref{eq:18}) and ~(\ref{eq:19}) yields
$C=\sqrt{2}\pi^{2}H \mbox{e}^{-\Phi}$, where $H$ and $\Phi$ are
calculated
at the first horizon-crossing and then
\begin{equation}
A_{h}=\frac{1}{\sqrt{2}} H \mbox{e}^{-\Phi} \;, \label{eq:27}
\end{equation}
which means that the amplitude of gravitational waves  produced during
inflation
directly depends on the given $f(R)$ theory since
$\Phi=\frac{1}{2}\ln f'(R)$. Explicitly, it is \cite{SCF1}
\begin{equation}
A_{h}=\frac{H}{\sqrt{2f'(R)}} \; ,\label{eq:28}
\end{equation}
where $f'(R)>0$ is necessary in order for the graviton to carry positive
kinetic energy \cite{review}.
The representation of $f(R)$ gravity as a Brans-Dicke theory is
particularly useful when  dealing with the scalar component of
gravitational waves,
ruled by the equation
\cite{SCF2}
\begin{equation}
\square \Phi=m^{2}\Phi \;,
\end{equation}
where $\Phi\equiv -\delta\phi/\phi_{0}$. The scalar field
generates a third component for
the tensor polarization of gravitational waves and the
total perturbation describing a  gravitational wave propagating in the
positive
$z$ direction is
\begin{eqnarray}
&& h_{\mu\nu}(t-z)=A^{+}(t-z)
\, e_{\mu\nu}^{(+)} +
A^{\times}(t-z) \, e_{\mu\nu}^{(\times)} \nonumber
\\
&&\nonumber \\
&& +\Phi(t-z) \, e_{\mu\nu}^{(s)} \;.  \label{eq:perturbazione totale}
\end{eqnarray}
The term
$A^{+}(t-z)e_{\mu\nu}^{(+)}+A^{\times}(t-z)e_{\mu\nu}^{(\times)}$
describes the two standard ({\em i.e.}, tensorial) polarizations
of a gravitational wave arising from GR in the TT
gauge \cite{Misner}, while the term $\Phi(t-z)e_{\mu\nu}^{(s)}$ is
the extension of the TT gauge mode to the scalar case. Three
different  degrees of freedom are present (see~eq.(32) of
\cite{capozzcorda}), while only two are present in standard
GR. Then,  for a purely scalar gravitational wave,
the metric perturbation is \cite{SCF2}
 \begin{equation}
h_{\mu\nu}=\Phi \, e_{\mu\nu}^{(s)} \;. \label{eq: perturbazionescalare}
\end{equation}
 The stochastic background of scalar gravitational waves can be
described in terms of the scalar field $\Phi$ and characterized by
a dimensionless spectrum (see the analogous definition for
tensor modes in
\cite{Allen,AO,Maggiore,Grishchuk})
\begin{equation}
\Omega_{sgw}(f)=\frac{1}{\rho_{c}}\frac{d\rho_{sgw}}{d\ln
f} \;,\label{eq:spettro}
\end{equation} where
 \begin{equation}
\rho_{c}\equiv\frac{3H_{0}^{2}}{8\pi G}\label{eq: densita'
critica}\end{equation}
is the (present) critical energy density of
the universe, $H_0$ is the  Hubble parameter today, and
$d\rho_{sgw}$ is the energy density of the scalar
gravitational radiation in the frequency interval $ \left( f,
f+df \right)$. We are now using standard units.
Now it is possible to write an
expression for the energy density of the stochastic scalar relic
gravitons background in the angular frequency interval
$(\omega,\omega+d\omega)$ as
\begin{equation}
d\rho_{sgw}=2\hbar\omega\left(\frac{\omega^{2}d\omega}{2\pi^{2}c^{3}}\right)N_{\omega}=
\frac{\hbar
H_{dS}^{2}H_{0}^{2}}{4\pi^{2}c^{3}}\frac{d\omega}{\omega}=\frac{\hbar
H_{dS}^{2}H_{0}^{2}}{4\pi^{2}c^{3}} \frac{df}{f}\,,\label{eq: de
energia}\end{equation}
where $f$, as above, is the frequency in
standard comoving time. Eq.~(\ref{eq: de energia}) can be
rewritten in terms of the critical and de Sitter  energy
densities
\begin{equation}
H_{0}^2 =\frac{8\pi G\rho_{c}}{3c^{2}}\,,\qquad
H_{dS}=\frac{8\pi G\rho_{dS}}{3c^{2}} \;.
\end{equation}
Introducing the Planck density ${\displaystyle
\rho_{Planck}=\frac{c^{5}}{\hbar G^{2}}}$, the spectrum is given
by
\begin{equation}
\Omega_{sgw}(f)=\frac{1}{\rho_{c}}\frac{d\rho_{sgw}}{d\ln
f}=\frac{f}{\rho_{c}}\frac{d\rho_{sgw}}{df}=\frac{16}{9}
\frac{\rho_{dS}}{\rho_{Planck}} \;.\label{eq:spettrogravitoni}
\end{equation}
At this point,  some  comments
are in order. First, the calculation works for a
simplified model that does not include the matter-dominated era.
If the latter is included, the redshift at the equivalence
epoch has to be considered. Taking into account Ref.~\cite{Allen2}, one
gets
\begin{equation}
\Omega_{sgw}(f)=\frac{16}{9}\frac{\rho_{dS}}{\rho_{Planck}}(
1+z_{eq})^{-1}
\label{eq:spettrogravitoniredshiftato}
\end{equation}
for the waves which,
at the epoch in which the universe becomes matter-dominated, have
a frequency higher than $H_{eq}$, the Hubble parameter at
equivalence. This situation corresponds to frequencies
$f>(1+z_{eq})^{1/2}H_{0}$ today. The redshift correction in
eq.~(\ref{eq:spettrogravitoniredshiftato}) is needed since the present value of
the
Hubble parameter $H_{0}$ would be different  without a matter-dominated
contribution. At lower frequencies, the spectrum is
given by \cite{Allen,Grishchuk}
\begin{equation}
\Omega_{sgw}(f)\propto f^{-2}.\label{eq:spettrobassefrequenze}
\end{equation} As a further consideration, let us note
that the results (\ref{eq:spettrogravitoni}) and
(\ref{eq:spettrogravitoniredshiftato}), which are frequency-independent, do not
hold in the entire range of physical
frequencies. For waves with frequencies less than the present Hubble parameter
$H_{0}$, the notion of energy density is not defined because the
wavelength becomes longer than the Hubble scale.
Similarly, at high frequencies, there is a maximal
frequency above which the spectrum rapidly drops to zero. In the
above calculation, the simplifying assumption that the phase transition
from the inflationary to the radiation dominated epoch is
instantaneous has been made. In the physical universe, this
process occurs over some time scale $\Delta\tau$, with
\begin{equation}
f_{max}=\frac{a(t_{1})}{a(t_{0})}\frac{1}{\Delta\tau} \;,
\label{eq:freq. max}\end{equation}
which is the redshifted rate of the
transition. In any case, $\Omega_{sgw}$ drops rapidly. The two
cutoffs at low and high frequencies for the spectrum guarantee
that the total energy density of the relic scalar gravitons is
finite. For GUT-scale inflation, it is of the order
\cite{Allen}
\begin{equation}
\frac{\rho_{ds}}{\rho_{Planck}}\approx10^{-12} \;.
\label{eq: rapportodensita' primordiali}
\end{equation}
These results can be
quantitatively constrained considering the recent {\em WMAP} release. In
fact, it is well known that {\em WMAP} observations put  severe
restrictions on the spectrum. In Fig.~\ref{fig:spectrum} the
spectrum $\Omega_{sgw}$ is mapped: considering the ratio
$\rho_{ds}/\rho_{Planck}$, the relic scalar gravitational wave  spectrum
seems
consistent with the {\em WMAP} constraints on scalar perturbations.
Nevertheless, since the spectrum falls off as $ f^{-2}$ at low
frequencies, today at {\em LIGO/VIRGO} and {\em LISA}
frequencies (indicated in Fig.~\ref{fig:spectrum}), one gets
\begin{equation} \Omega_{sgw}(f)h_{100}^{2}<2.3\times
10^{-12} \;,\label{eq:limitespettroWMAP}
\end{equation}
where $h_{100}=H_0/\left( 100 \; \mbox{km}\cdot \mbox{s}^{-1}
\cdot \mbox{Mpc}^{-1} \right)$.  It is interesting to calculate
the  corresponding strain at $ f \sim 100$Hz, where
interferometers such as {\em VIRGO} and {\em LIGO} achieve maximum
sensitivity. The well known equation for the characteristic
amplitude \cite{Allen,Grishchuk} adapted to the scalar component
of gravitational waves
\begin{equation}
\Phi_{c}(f)\simeq1.26\times
10^{-18}\left(
\frac{1 \, \mbox{Hz}}{f} \right)\sqrt{h_{100}^{2}\Omega_{sgw}(f)}
\;, \label{eq:legameampiezza-spettro}
\end{equation}
can be used to obtain
\begin{equation}
\Phi_{c}\left( 100\, \mbox{Hz} \right) < 2\cdot 10^{-26}\;.
\label{eq:limiteperlostrain}
\end{equation}

Then, since we expect a sensitivity of the order of $10^{-22}$ for
the above interferometers at $f \sim 100$~Hz, we need to gain four
orders of magnitude. Let us analyze the situation also at lower
frequencies. The sensitivity of the {\em VIRGO} interferometer is of the
order of $10^{-21}$ at $f\sim 10$~Hz and in that case it is
\begin{equation}
\Phi_{c} \left( 10 \, \mbox{Hz} \right) < 2\cdot 10^{-25} \;.
\label{eq:limiteperlostrain2}\end{equation}
The sensitivity of the
{\em LISA} interferometer
will be of the order of $10^{-22}$ at $ f \sim 10^{-3} $~Hz and in
this case it is
\begin{equation}
\Phi_{c} \left( 10^{-3} \mbox{Hz} \right) <2\cdot 10^{-21} \;.
\label{eq:limiteperlostrain3}
\end{equation}
This means that a stochastic background of
relic scalar gravitational waves could, in  principle, be detected by
the {\em LISA} interferometer.

To estimate the rms deflection effect from massive scalar modes in
the gravitational wave background, we restrict to periods that are
of the order of hours or days. A longer period would result in a
``frozen'' effect which is much less likely to be detected, while
a much shorter period would probably render the deflections
unobservable because only an averaged position shift would be
recorded during observation times longer than the period itself
and a slightly blurred image would be the outcome (although fast
photometry might allow to push the limits).  Assuming a frequency
$\sim 10^{-3}$~Hz at a distance $L\sim 500 $~kpc and using the
upper limit~(\ref{eq:limiteperlostrain3}), one obtains a rms
deflection $\theta_{rms} \sim \sqrt{kL}\, \epsilon \sim 10^{-10}$.
The maximum resolution expected with high precision astrometry is
of the order of microarcseconds ($\sim 10^{-7}$ radians), three
orders of magnitude above the required sensitivity for detection.
For galactic sources at $L\sim 5 $~kpc, to which high precision
astrometry is more likely to apply, $\theta_{rms}$ drops by
another order of magnitude. One can, of course, consider different
sources of electromagnetic radiation with slightly higher
frequency, at more promising distances $L$, and perhaps find
mechanism which produce higher scalar amplitudes $\Phi$: at a
first look, however, it is unlikely that the four orders of
magnitude necessary for detection can be bridged in the
foreseeable future.

\section{V. Outlooks}

All modern theories of high energy physics unifying gravity with
the other interactions predict departures from GR;
however, no such deviation has been observed so far in Solar
System experiments, and practically all the experimental
constraints on such deviations are obtained within the Solar
System (the binary pulsar and gravitational lensing provide
constraints that are not competitive with those obtained from
Solar System experiments). Therefore, it is interesting to
explore astrophysical
effects outside of this narrow region of the universe that could
potentially exhibit deviations from Einstein's theory. The
cumulative deflections, or frequency shifts due to propagation of
light from distant sources through random massive modes of the
gravitational wave background could constitute such an effect.
There is now a wide range of theories predicting massive scalar,
vector, and tensor modes that can lead to such an effect. However,
the astrophysical and cosmological processes generating
cosmological gravitational wave backgrounds in these theories are
still unexplored. Here we do not attempt to estimate the average
strength $\epsilon$  of the  various modes appearing in these
theories of gravity, in different ranges of wavelengths, and under
various assumptions. We limit ourselves to  outline an estimate
for  a particular case: massive scalar modes in $f(R)$ gravity.
This  class of theories has been the subject of much recent
attention  in order to explain the observed acceleration of the
universe without resorting to dark energy. For the situation
considered here, the deflections seem to be a few orders of
magnitude too small for detection; on the other hand, we find that
these scalar modes are certainly of interest for direct attempts
at detection with the {\em LISA} experiment.  According to our
preliminary discussion, the indirect  detection with the position
and frequency shift effect does not seem to be feasible with
current technology; however, a more detailed analysis is necessary
before definitive conclusions can be drawn. It is, in principle,
possible that massive gravitational wave modes could be produced
in more significant quantities in cosmological or early
astrophysical processes in alternative theories of gravity---the
latter are still unexplored. This possibility should be kept in
mind when looking for a signature distinguishing these theories
from GR, and seems to deserve further
investigation.

\begin{acknowledgments}
VF acknowledges the Natural
Sciences and  Engineering Research Council of Canada for financial support and INFN-Laboratori Nazionali di
Frascati, the University of Naples ``Federico~II'', and  the
International School for Advanced Studies in Trieste
for their hospitality.
\end{acknowledgments}


\end{document}